\documentclass[traditabstract,printer]{aa}
\usepackage{graphicx}
\usepackage{epsfig}
\usepackage{amssymb,amsmath}
\usepackage{natbib}
\bibpunct{(}{)}{;}{a}{}{,}
\usepackage{pslatex}
\usepackage{caption}
\usepackage{longtable}
\usepackage{hyperref}
\usepackage{gensymb}

\begin{document}

\title{Redshift evolution of the Amati relation: calibrated results from the Hubble diagram of quasars at high redshifts}

\author{Yan Dai$^{1,2}$, Xiao-Gang Zheng$^{3}$, Zheng-Xiang Li$^{1}$\thanks{E-mail: \href{zxli918@bnu.edu.cn}{zxli918@bnu.edu.cn}}, He Gao$^{1}$\thanks{E-mail: \href{gaohe@bnu.edu.cn}{gaohe@bnu.edu.cn}} \& Zong-Hong Zhu$^{1}$\thanks{E-mail: \href{zhuzh@bnu.edu.cn}{zhuzh@bnu.edu.cn}}}

\institute{$^1$Department of Astronomy, Beijing Normal University, Beijing 100875, China;\\
$^2$Beijing Planetarium, Beijing 100044, China;\\
$^3$School of Electrical and Electronic Engineering, Wuhan Polytechnic University, Wuhan 430023, China\\
}

\authorrunning{Dai}
\titlerunning{Redshift evolution of the Amati relation}

\abstract{Gamma-ray bursts (GRBs) have long been proposed as a complementary probe to type Ia supernovae (SNe Ia) and cosmic microwave background to explore the expansion history of the high-redshift universe, mainly because they are bright enough to be detected at greater distances. Although they lack definite physical explanations, many empirical correlations between GRB isotropic energy/luminosity and some directly detectable spectral/temporal properties have been proposed to make GRBs standard candles. Since the observed GRB rate falls off rapidly at low redshifts, thus preventing a cosmology independent calibration of these correlations. In order to avoid the circularity problem, SN Ia data are usually used to calibrate the luminosity relations of GRBs in the low redshift region (limited by the redshift range for SN Ia sample), and then extrapolate it to the high redshift region. This approach is based on the assumption of no redshift evolution for GRB luminosity relations. In this work, we suggest the use a complete quasar sample in the redshift range of $0.5<z<5.5$ to test such an assumption. We divide the quasar sample into several sub-samples with different redshift bins, and use each sub-sample to calibrate the isotropic $\gamma$-ray equivalent energy of GRBs in relevant redshift bins. By fitting the newly calibrated data, we find strong evidence that the most commonly used \emph{Amati} relation between spectral peak energy and isotropic-equivalent radiated energy shows no, or marginal, evolution with redshift. Indeed, at different redshifts, the coefficients in the \emph{Amati} relation could have a maximum variation of 0.93\% at different redshifts, and there could be no coincidence in the range of 1$\sigma$. 
}

%$\log \left({E_{p,z}}/{\rm keV}\right)=a\log \left({E_{\rm iso}}/{\rm erg}\right)+b$, 


\maketitle

\section{Introduction}

As the most intense explosions in the universe, gamma-ray bursts (GRBs) are bright enough to be detected in high-redshift range up to at least $z\sim10$ \citep{tanvir09,salvaterra09}, so that GRBs have been widely discussed as a complementary probe to SNe Ia and cosmic microwave background (CMB) to explore the expansion history of the high-redshift universe \cite[][for a review]{amati13,wang15}. In order to make GRBs as standard candles, many empirical correlations between their isotropic energy/luminosity and some directly detectable spectral/temporal properties have been proposed. For instance, \cite{amati02} discovered a correlation between the isotropic bolometric emission energy ($E_{\rm iso}$) and the rest-frame peak energy ($E_{\rm p,z}$), later \cite{ghirlanda04a} proposed that there exists an even tighter correlation between $E_{\rm p,z}$ and the beaming-corrected bolometric emission energy ($E_{\gamma}$). In order to reduce the scatter of the correlations, several multiple relations have also been proposed, such as the $E_{\rm iso}$ - $E_{\rm p,z}$ - $t_{\rm b,z}$ relation \citep{liang05} and so on \cite[see more correlations reviewed in][]{ghirlanda04b,demianski17a,zhang18}. 

These GRB luminosity indicators have been widely used as standard candles for cosmology research \citep{schaefer07,wang07,amati08,amati13,wei13,wang17,amati19,muccino21,montiel21}, and showed that when combined with other probes GRBs can indeed extend the Hubble diagram to higher redshifts and help to make better constraints on cosmological parameters \cite[][for a review]{amati13,wang15}. However, since the observed GRB rate falls off rapidly at low redshifts, it is very difficult to make a robust calibration for those GRB correlations. Normally, a robust cosmology independent calibration (e.g. the standard $\Lambda$CDM cosmology) would be used to calculate the isotropic energy/luminosity for GRB samples and then derive the empirical correlations. As a result, the so-called circularity problem could prevent the direct use of GRBs for cosmology. 

To avoid the circularity problem, it has been proposed to use SN Ia data in the same redshift range as GRBs to calibrate their luminosity correlations \citep{liang08,vitagliano10,gao12}. Considering that objects at the same redshift should have the same luminosity distance in any cosmology, one can assign the distance moduli of SN Ia to GRBs at the same redshifts \citep{liang08,demianski17a}, or one can use the SN Ia data to fit the model-independent cosmography formula that reflects the Hubble relation between luminosity distance and redshift, and then obtain the GRB luminosity distance \citep{gao12,amati19}. In this way, one can derive a cosmology-independent calibration to the GRB candles. 

The shortage of using SN Ia to calibrate GRB correlations is that the redshift range of SN Ia data is relatively low. In this case, one can only calibrate the correlations for low redshift GRB sample and extend the calibrated relations to high redshift. For this method, one needs to make a hypothesis that there is no evolution with respect to redshift for the GRB correlations. It is of great interest and necessary to test this hypothesis because the high redshift GRB samples are the most important for the study of cosmology. 

Recently, based on a non-linear relation between quasars' UV and X-ray luminosities [parametrized as $\log(L_{X})=\gamma\log(L_{\rm UV})+\beta$], \cite{risaliti19} have constructed a Hubble diagram of quasars in redshift range of $0.5<z<5.5$, which is in excellent agreement with the analogous Hubble diagram for SNIa in the redshift range of $0.5<z<1.4$. Here we suggest to divide the quasar sample into several sub-samples with different redshift bins, and use each sample to calibrate GRB correlations to test if there exists redshift evolution for these relations.

\section{Amati relation and GRB sample}

Although, many empirical luminosity correlations have been statistically found from long GRB observations, the Amati correlation is the most widely used one for cosmological studies. In this work, we focus on the discussion for the redshift dependence of Amati relation. This relation was first found for a sample of long GRBs detected by BeppoSAX with known redshifts \citep{amati02,amati06}, showing that more energetic GRBs tend to be spectrally harder. With the increase of long GRB events, Amati relation always holds, although a few significant outliers do exist \citep{amati08}. Thanks to the successful operation of the \emph{Neil Gehrels Swift Observatory} \citep{gehrels04}, a good sample of short GRBs were well localized, whose redshift were precisely measured. It is found that short GRBs also have the correlation between the isotropic bolometric emission energy and the rest-frame peak energy, but they seem to form a parallel track above the long-GRB Amati relation \citep{ghirlanda09}. For this work, we only focus on the long GRB sample.

The Amati relation has the form as 
\begin{eqnarray}
\frac{E_{\rm p,z}}{100{\rm keV}}=C\left(\frac{E_{\rm iso}}{10^{52}{\rm erg}}\right)^m
\end{eqnarray}
with $C\sim0.8-1$ and $m\sim0.4-0.6$ \citep{amati06}. $E_{\rm iso}$ is the isotropic equivalent energy in gamma-ray band, which can be calculated from the bolometric fluences $S_{\rm {\bf{bol}}}$~as 
\begin{equation}\label{equ:Eiso}
E_{\rm iso}=4\pi d^{2}_{L}S_{\rm {\bf{bol}}}\left(1+z \right)^{-1}~,
\end{equation}
where $S_{\rm {\bf{bol}}}$ is calculated from the observed fluence in the rest frame $1-10,000$ keV energy band by assuming the Band function spectrum \citep{band93}
\begin{equation}
  N(E) = \left\{ \begin{array}{ll}
    A \left(\frac{E}{100{\rm keV}}\right)^{\alpha}{\rm exp}\left(
    \frac{E}{E_0} \right), & E < (\alpha-\beta)E_0\\
    A \left(\frac{(\alpha-\beta)E_0}{100{\rm keV}}\right)^{\alpha}{\rm exp}\left(\alpha-\beta
     \right)\left(\frac{E}{100{\rm keV}}\right)^{\beta}, & E > (\alpha-\beta)E_0
  \end{array} \right.
\end{equation}
$E_{\rm p,z}=(1+z)E_{p}$ is the rest-frame peak energy, where $E_p=(\alpha+2)E_0$ is the peak energy in the $E^2N(E)$ spectrum. 

For the purpose of this work, we express the Amati relation as 
\begin{eqnarray}
%\log\left(\frac{E_{p,z}}{{\rm keV}}\right)=a\log\left(\frac{E_{\rm iso}}{{\rm erg}}\right)+b.
\log \left(\frac{E_{{\bf{\rm{p,z}}}}}{\rm keV}\right)=a\log \left(\frac{E_{\rm iso}}{\rm erg}\right)+b.
\end{eqnarray}
Here we adopt the GRB sample compiled in~\cite{demianski17a,demianski17b}, which includes 162 well-measured GRB in the redshift range: $z\in [0.125,~9.3]$. We divide the GRB sample into four sub-groups with different redshift bins, e.g., $0.125<z\le1$, $1<z\le2$, $2<z\le3$ and $z>3$. The number of GRBs contained in the 4 subsamples is 42, 54, 35 and 30, respectively. 

\begin{figure}\label{fig1}
\centering
\includegraphics[width=10cm,height=7.5cm]{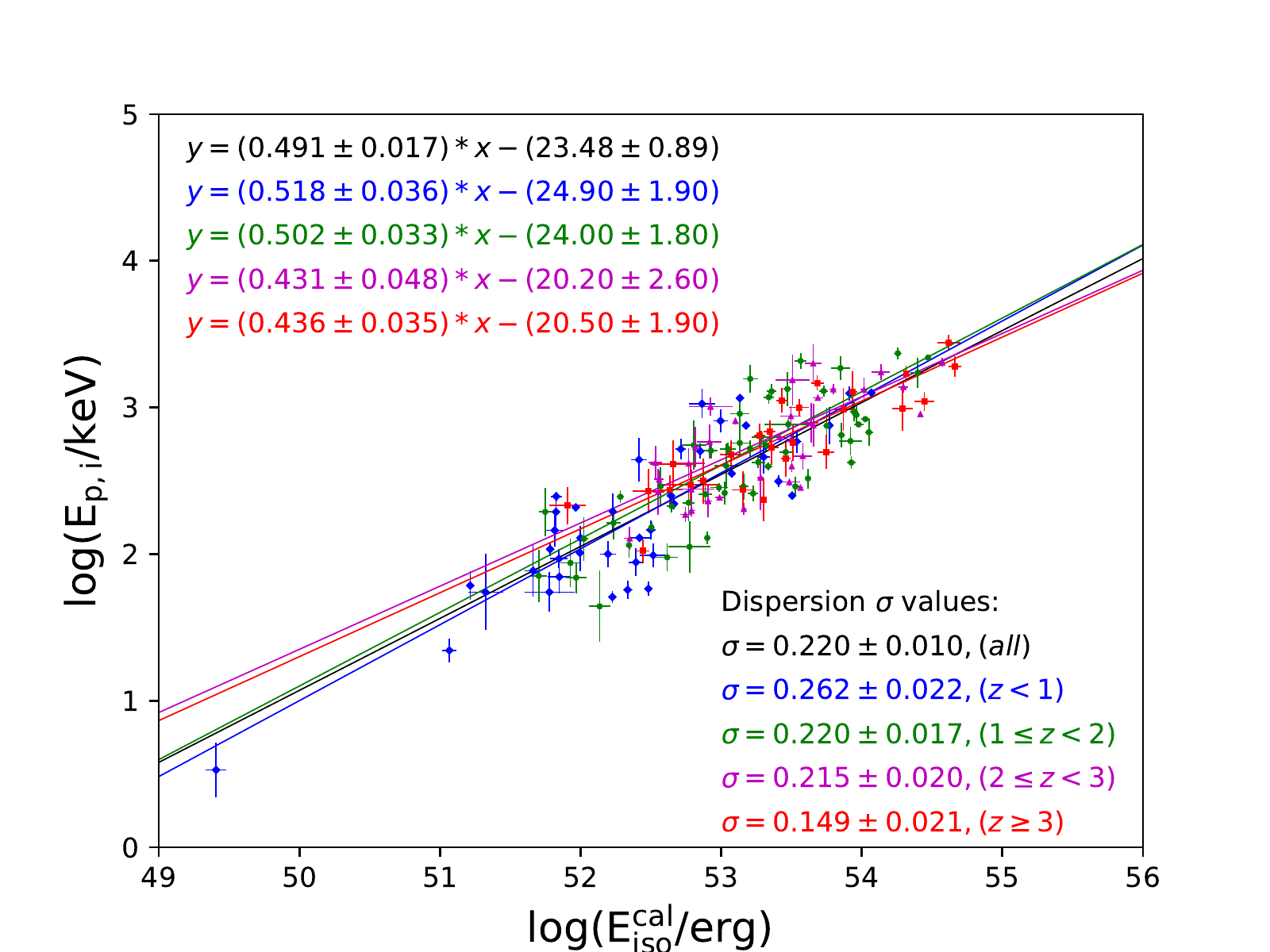}\\
\includegraphics[width=8cm,height=7.5cm]{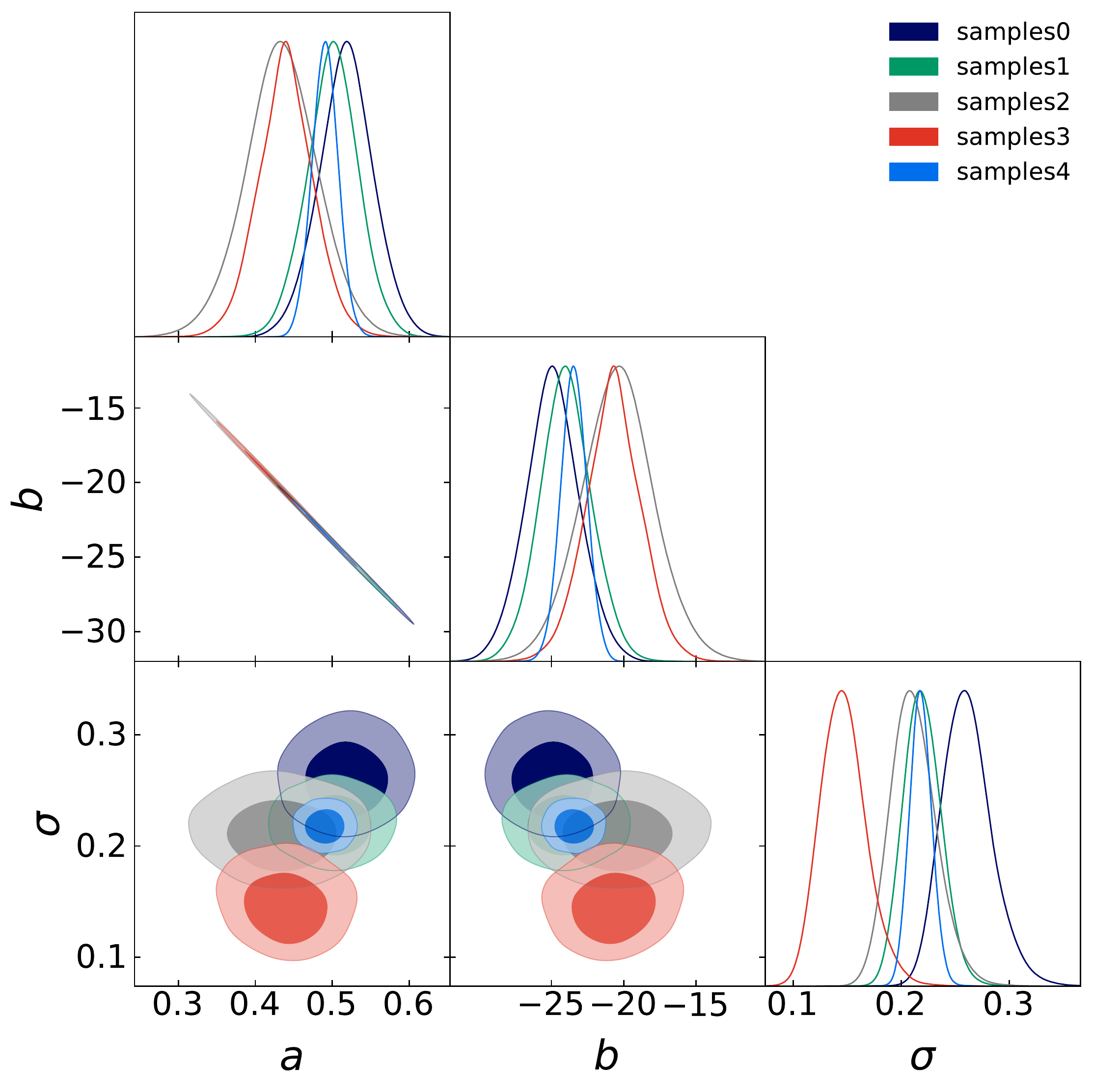}
\caption{Upper panel: fitting results for Amati relation with GRB samples in different redshift range (black line is for the whole sample, blue line is for $0.125<z\le1$, green line is for $1<z\le2$, purple line is for $2<z\le3$ and red line is for $z\ge3$), where the GRB distance moduli are calibrated by the quasar data. Lower panel: the 1-D marginalized distributions and 2-D plots with $1\sigma$ and $2\sigma$ contours for luminosity correlation parameters.} 
\end{figure}

\begin{table}
\begin{center}{\scriptsize
\caption{Best fitting values of coefficient $a$ (upper panel) and $b$ (lower panel) for each sub-sample}
\begin{tabular}{cccc}
\hline
\hline
& $a$ & $b$ & $\delta$\\
\hline
$0.125<z\le1$ & $0.518\pm0.036$ & $24.90\pm1.90$ & $0.262\pm0.022$\\
$1<z\le2$ & $0.502\pm0.033$ & $24.00\pm1.80$ & $0.220\pm0.017$\\
$2<z\le3$ & $0.431\pm0.048$ & $20.20\pm2.60$ & $0.215\pm0.020$\\
$z>3$ & $0.436\pm0.035$ & $20.50\pm1.90$ & $0.149\pm0.021$\\
Total & $0.491\pm0.017$ & $23.48\pm0.89$ & $0.220\pm0.010$\\
\hline
 \end{tabular}
 }
\end{center}
\end{table}

\section{Calibration results for Amati relation using Hubble diagram of Quasars}

As the most luminous persistent sources in the Universe, quasars are bright enough to be detected up to redshifts $z>7$ \citep{mortlock11,banados18,wang18,yang20}. According to the currently accepted model, quasars are extremely luminous active galactic nucleus (AGN), where the observed intense energy release are related to the accretion of a gaseous disk onto a supermassive black hole (SMBH). Quasars have a wide spectral energy distribution, which normally contains a significant emission component in the optical-UV band $L_{UV}$, the so-called big blue bump, with a softening at higher energies \citep{sanders89,elvis94,trammell07,shang11}. It has long been discussed that there is a non-linear relationship between $L_{UV}$ and the quasar's X-ray luminosity $L_X$, parametrized as $\log(L_{X})=\gamma\log(L_{\rm UV})+\beta$ \citep{vignali03,strateva05,steffen06,just07,green09,young10,jin12}. From the theoretical point of view, this relation could be intrinsic, since the UV emission is usually thought to origin from the optically thick disc surrounding the SMBH and the X-ray photons are thought to be generated through inverse-Compton scattering of these disk UV photons by a plasma of hot relativistic electrons (the so-called corona) around the accretion disk. Such relation is found to be independent of redshift \citep{lusso16}, so that it could be used as a distance indicator to estimate cosmological parameters. The initial dispersion of the $L_{UV}-L_X$ relation is relatively large ($\delta\sim0.35-0.4$, \citep{just07,young10}), but after a detailed study, \cite{lusso16} suggest that most of the observed dispersion is not intrinsic, but it is rather due to observational effects. By gradually refining the selection technique and flux measurements, \cite{risaliti19} have collected a complete sample of quasars, whose dispersion of $L_{UV}-L_X$ relation is smaller than 0.15 dex. The main quasars sample is composed of 1598 data points in the range $0.036<z<5.1$.  With this sample, they constructed a Hubble diagram of quasars in redshift range of $0.5<z<5.5$, which is in excellent agreement with the analogous Hubble diagram for SNIa in the redshift range of $0.5<z<1.4$. Moreover, this Hubble diagram of quasars have been studied in cosmological applications \citep{zheng20,zheng21}. Considering that objects at the same redshift should have the same luminosity distance in any cosmology, here we first fit the model-independent cosmography formula that reflects the Hubble relation between luminosity distance and redshif using the quasar sample, and then obtain the distance moduli (also the luminosity distance) for GRBs at a given redshift with the best fit results. 

It has long been proposed that the evolution of the universe could be described by pure kinematics, only relying on the assumption of the basic symmetry principles (the cosmological principle) that the universe can be described by the Friedmann-Robertson-Walker metric, but independent of any cosmology model \citep{weinberg72}. In such a cosmography framework, the luminosity distance could be expressed as a power series in the redshift by means of a Taylor series expansion \citep{visser04}
\begin{eqnarray}
d_{L}(z)=&&cH_0^{-1}\left\{z+\frac{1}{2}(1-q_0)z^2-\frac{1}{6}\left(1-q_0-3q_0^2+j_0+\right.\right.\nonumber
  \\ &&\left.\left.+\frac{kc^2}{H_0^2a^2(t_0)}\right)z^3+\frac{1}{24}\left[2-2q_0-15q_0^2-15q_0^3+5j_0+\right.\right.\nonumber
  \\ &&\left.\left.+10q_0j_0+s_0+\frac{2kc^2(1+3q_0)}{H_0^2a^2(t_0)}\right]z^4+...\right\}\, ,
\end{eqnarray}
where $k=-1,0,+1$ corresponds to hyperspherical, Euclidean or spherical universe, respectively. The coefficients of the expansion are the so-called cosmographic parameters (e.g. Hubble parameters $H$, deceleration parameters $q$, jerk parameters $j$, and snap parameters $s$), which relate to the scale factor $a(t)$ and its higher order derivatives :
\begin{eqnarray}
H&\equiv&\frac{\dot{a}(t)}{a(t)}~,\label{eq:H}\\
q&\equiv&-\frac{1}{H^2}\frac{\ddot{a}(t)}{a(t)}~,\label{eq:q}\\
j&\equiv&\frac{1}{H^3}\frac{a^{(3)}(t)}{a(t)}~,\label{eq:j}\\
s&\equiv&\frac{1}{H^4}\frac{a^{(4)}(t)}{a(t)}~.\label{eq:s}
\end{eqnarray}
All subscripts ``0'' indicate the present value of the parameters ($t=t_0$). In order to avoid the convergence of the series at high redshift, and to better control the approximation induced by truncations of the expansions, \cite{cattoen07} proposed to use an improved parameter $y=z/(1+z)$ to recast $d_L$ expression as 
\begin{eqnarray}
d_L(y)=&&\frac{c}{H_0}\left\{y-\frac{1}{2}(q_0-3)y^2+\frac{1}{6}\left[12-5q_0+3q^2_0-\right.\right.\nonumber
  \\ &&\left.\left.-(j_0+\Omega_0)\right]y^3+\frac{1}{24}\left[60-7j_0-10\Omega_0-32q_0+\right.\right.\nonumber
  \\ &&\left.\left.+10q_0j_0+6q_0\Omega_0+21q^2_0-15q^3_0+s_0\right]y^4+\mathcal{O}(y^5)\right\}~,
\end{eqnarray}
where $\Omega_0$ is the total energy density. For the purpose of this work, here we fit the truncation of the $d_L$ expression (for the flat universe $\Omega_0=1$) to the second order term with the quasar sample. Our best cosmographic fitting result is 
\begin{equation}
q_0=-0.64\pm0.61~~,~~~j_0+\Omega_0=0.92\pm5.04\nonumber~. 
\end{equation}
In our adopted sample, 156 GRBs are in the redshift range of $0.5<z<5.5$. Luminosity distance for these GRBs are re-calculated based on the best fitted cosmography formula, so are for their isotropic equivalent energy in gamma-ray band.

\begin{figure}
\centering
\includegraphics[width=9.5cm,height=7.5cm]{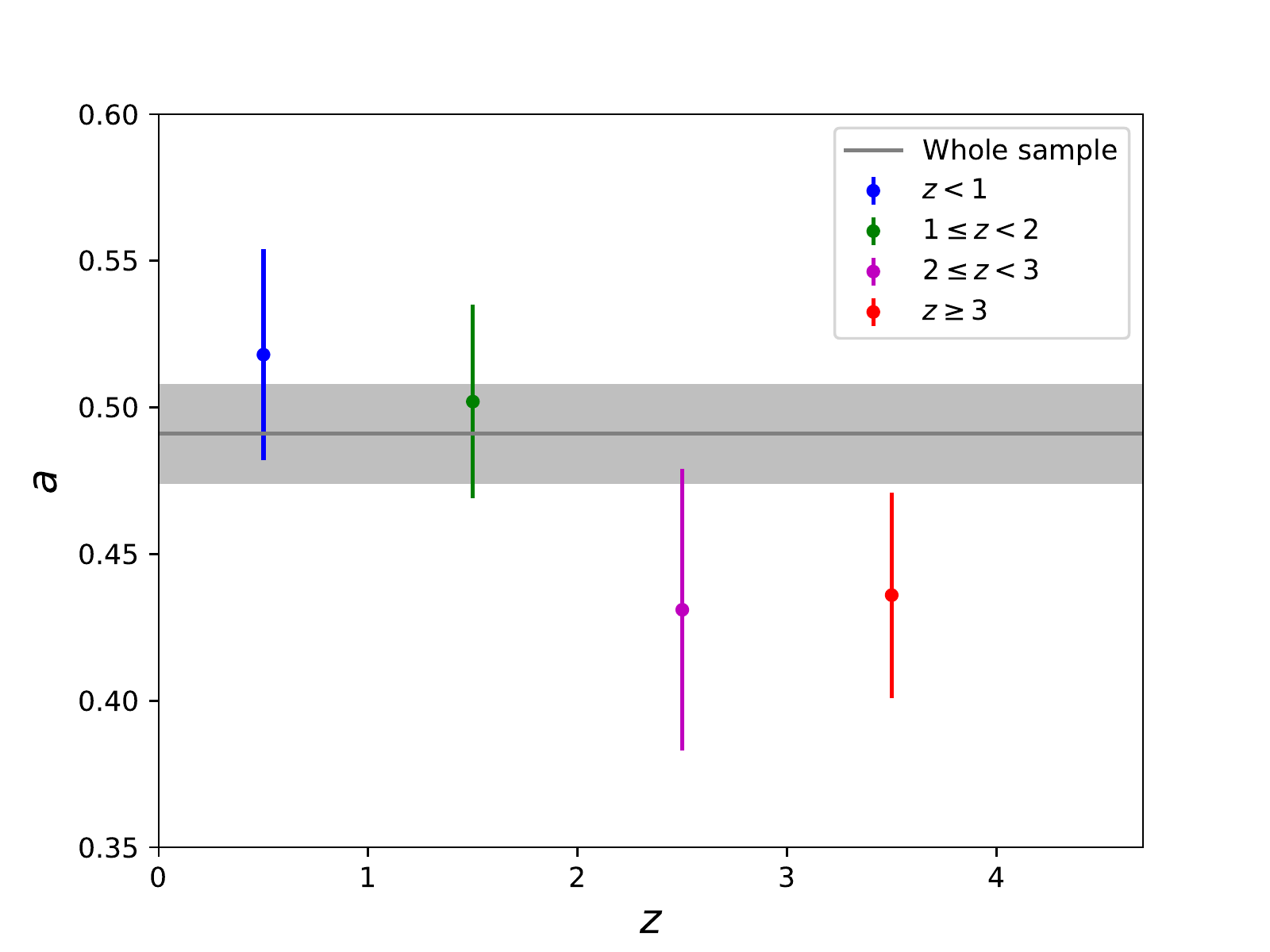} 
\includegraphics[width=9.5cm,height=7.5cm]{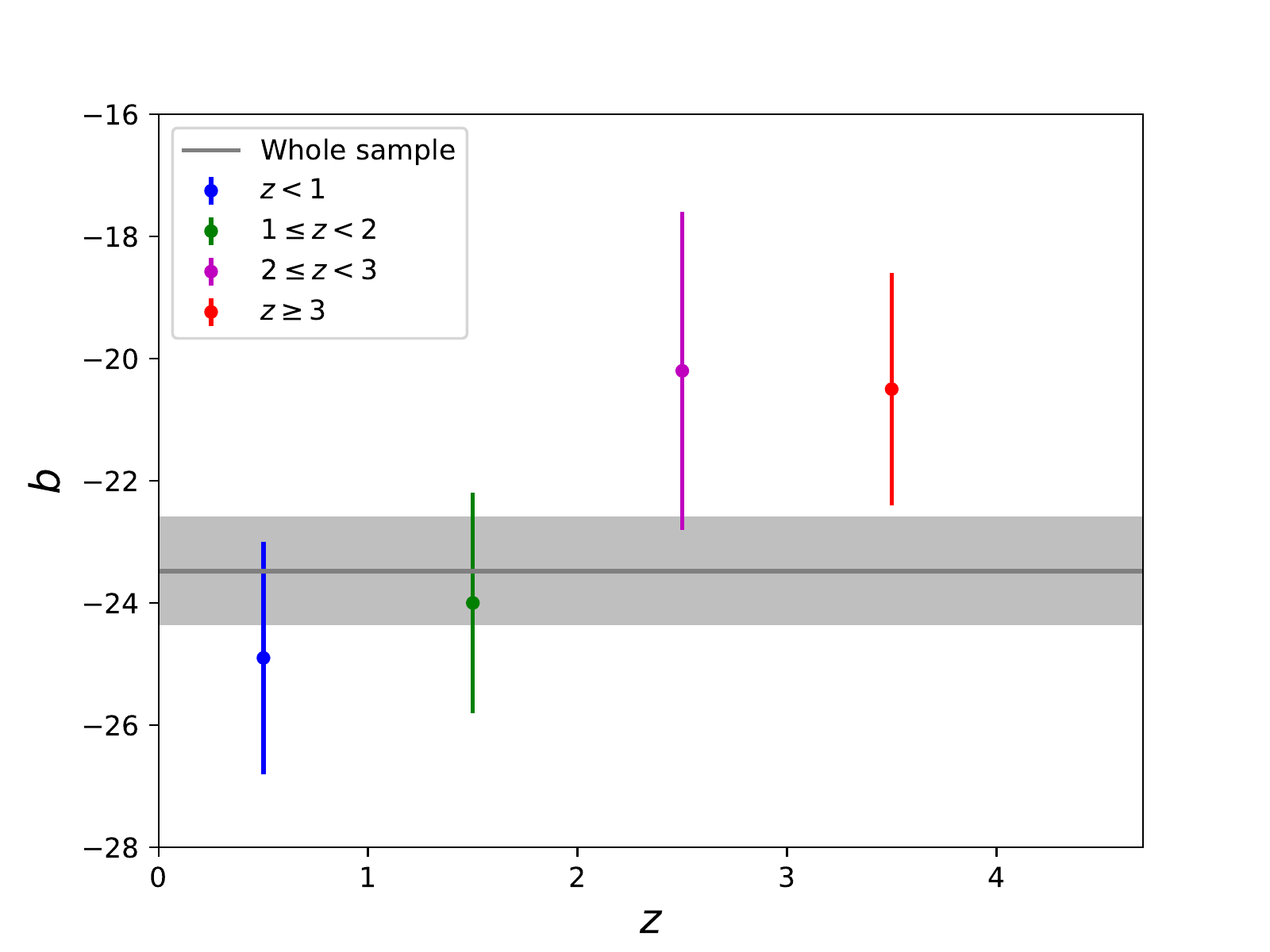} 
\caption{Best fitting values of coefficient $a$ (upper panel) and $b$ (lower panel) for each sub-sample with respect to the sub-sample's mean redshift.}
\end{figure}

With the newly calculated isotropic equivalent energy $E_{\rm iso}$, we calibrate the Amati relation for each sub-sample and the whole sample. Here we make a logarithm linear fitting between $E_{\rm p,z}$ (in unit of keV) and $E_{\rm iso}$ (in unit of ergs) by adopting a likelihood function written as \citep{Reichart01}

\begin{eqnarray}
L(a, b, \sigma) & = & \frac{1}{2} \frac{\sum{\log{(\sigma^2 + \sigma_{y_i}^2 + a^2
\sigma_{x_i}^2)}}}{\log{(1+a^2)}} \\ \nonumber &&+ \frac{1}{2} \sum{\frac{(y_i - a x_i - b)^2}{\sigma^2 + \sigma_{y_i}^2 + a^2
\sigma_{x_i}^2}}\,,\label{eq: deflike}
\end{eqnarray}
where $x_i=\log (E_{\rm iso})$, $y_i=\log (E_{\rm p,z})$ and $\sigma$ marks the observational intrinsic dispersion. We adopt the Python package of \emph{emcee} to perform the fitting and take the Uniform priors on $a\in[0,4]$, $b\in[50, 56]$, and $\delta\in[0,1]$. The best fitting results for each sub-sample and the whole sample are collected in Table 1 and plotted in Figure 1. In the bottom panel of Figure 1 we show the 1-D marginalized distributions and 2-D contours with $1\sigma$ and $2\sigma$ confidence region. In Figure 2, we plot the best fitting values of coefficient $a$ and $b$ for each sub-sample with respect to the sub-sample's mean redshift. 

Our results show that the best fitting values of $a$ and $b$ seem to have a certain redshift evolution. For coefficient $a$, the results of the first three sub-samples ($z<3$) are in good agreement with each other, however the result of the last high redshift sub-sample does not coincide with the previous ones in the range of 1$\sigma$, but still coincide in the range of 2$\sigma$. More interestingly, the best fitting values of coefficient $b$ first increases and then decreases with the increase of sample redshift. The variation range of $b$ value between different sub-samples could reach $0.93\%$, and the result for medium redshift range ($1<z<3$) does not coincide with the low or high redshift sample in range of 1$\sigma$. As shown in the lower panel of Figure 1, parameters a and b are correlated, which is expected for a linear fit. There seems no evolution trend for the dispersion (denoted by the $\delta$ value) of the relation. 

\section{Discussion}

Gamma-ray bursts are attractive cosmic probes due to their high redshift characteristics. Many empirical correlations between their isotropic energy/luminosity and some directly detectable spectral/temporal properties have been proposed, intending to shape GRBs into standard candles. However, unlike the Type Ia SNe, all these GRB luminosity relations lack definite physical explanations, mainly because our knowledge of the progenitor, central engine and jet composition for GRBs are still limited. In this case, before applying GRBs to explore the universe, the properties of these relations need to be further examined. For example, whether there is redshift evolution for these relations is a problem worthy of study. Using a complete quasar sample with large redshift span to calibrate the isotropic equivalent energy of GRBs, here we find no significant evidence that the Amati relation has an evolution with redshift. Some previous methods, such as calibrating the luminosity relations of GRBs in the low redshift region, and then extrapolating it to the high redshift region, may be problematic. It is worth noticing that the dispersion of Hubble diagram for our adopted quasar sample is relatively large, which may bring some uncertainty to the distance calibration, so as to the fitting results for the Amati relation. But the method we discussed here is universal. In the future, when the quasar sample quality becomes better, or there are other better distance indicator samples with large redshift span, we will study the redshift evolution of GRB luminosity relation better.

\acknowledgements
This work was supported by the National Natural Science Foundation of China under Grants Nos. 11633001, 11920101003, 12021003 and 11690024, the Strategic Priority Research Program of the Chinese Academy of Sciences, Grant No. XDB23000000 and the Interdiscipline Research Funds of Beijing Normal University.

\label{lastpage}


\begin{thebibliography}{}

\bibitem[Amati et al.(2002)]{amati02} Amati, L., Frontera, F., Tavani, M., et al.\ 2002, \aap, 390, 81. doi:10.1051/0004-6361:20020722
\bibitem[Amati(2006)]{amati06} Amati, L.\ 2006, \mnras, 372, 233. doi:10.1111/j.1365-2966.2006.10840.x
\bibitem[Amati et al.(2008)]{amati08} Amati, L., Guidorzi, C., Frontera, F., et al.\ 2008, \mnras, 391, 577. doi:10.1111/j.1365-2966.2008.13943.x
\bibitem[Amati \& Della Valle(2013)]{amati13}  Amati, L. \& Della V.\ 2013, International Journal of Modern Physics D, 22, 14, 1330028. doi.org/10.1142/S0218271813300280
\bibitem[Amati et al.(2019)]{amati19} Amati, L., D'Agostino, R., Luongo, O., et al.\ 2019, \mnras, 486, L46. doi:10.1093/mnrasl/slz056
\bibitem[Banados et al. (2018)]{banados18} Banados, E., Venemans, B., Mazzucchelli, C., et al.\ 2018, \nat, 553, 473. doi:10.1038/nature25180
\bibitem[Band et al.(1993)]{band93} Band, D., Matteson, J., Ford, L., et al.\ 1993, \apj, 413, 281. doi:10.1086/172995
\bibitem[Catto{\"e}n \& Visser(2007)]{cattoen07} Catto{\"e}n, C. \& Visser, M.\ 2007, Classical and Quantum Gravity, 24, 5985. doi:10.1088/0264-9381/24/23/018
\bibitem[Demianski et al.(2017a)]{demianski17a} Demianski, M., Piedipalumbo, E., Sawant, D., et al.\ 2017a, \aap, 598, A112. doi:10.1051/0004-6361/201628909
\bibitem[Demianski et al.(2017b)]{demianski17b} Demianski, M., Piedipalumbo, E., Sawant, D., et al.\ 2017b, \aap, 598, A113. doi:10.1051/0004-6361/201628911
\bibitem[Elvis et al.(1994)]{elvis94} Elvis, M., Wilkes, B.~J., McDowell, J.~C., et al.\ 1994, \apjs, 95, 1. doi:10.1086/192093
\bibitem[Gao et al.(2012)]{gao12} Gao, H., Liang, N., \& Zhu, Z.-H.\ 2012, International Journal of Modern Physics D, 21, 1250016-1-1250016-16. doi:10.1142/S0218271812500162
\bibitem[Gehrels et al.(2004)]{gehrels04} Gehrels, N., Chincarini, G., Giommi, P., et al.\ 2004, \apj, 611, 1005. doi:10.1086/422091
\bibitem[Ghirlanda et al.(2004a)]{ghirlanda04a} Ghirlanda, G., Ghisellini, G., \& Lazzati, D.\ 2004, \apj, 616, 331. doi:10.1086/424913
\bibitem[Ghirlanda et al.(2004b)]{ghirlanda04b} Ghirlanda, G., Ghisellini, G., Lazzati, D., et al.\ 2004, \apjl, 613, L13. doi:10.1086/424915
\bibitem[Ghirlanda et al.(2009)]{ghirlanda09} Ghirlanda, G., Nava, L., Ghisellini, G., et al.\ 2009, \aap, 496, 585. doi:10.1051/0004-6361/200811209
\bibitem[Green et al.(2009)]{green09} Green, P.~J., Aldcroft, T.~L., Richards, G.~T., et al.\ 2009, \apj, 690, 644. doi:10.1088/0004-637X/690/1/644
\bibitem[Jin et al.(2012)]{jin12} Jin, C., Ward, M., \& Done, C.\ 2012, \mnras, 422, 3268. doi:10.1111/j.1365-2966.2012.20847.x
\bibitem[Just et al.(2007)]{just07} Just, D.~W., Brandt, W.~N., Shemmer, O., et al.\ 2007, \apj, 665, 1004. doi:10.1086/519990
\bibitem[Khadka \& Ratra(2020)]{khadka20} Khadka, N. \& Ratra, B.\ 2020, \mnras, 497, 263. doi:10.1093/mnras/staa1855
\bibitem[Liang \& Zhang(2005)]{liang05} Liang, E. \& Zhang, B.\ 2005, \apj, 633, 611. doi:10.1086/491594
\bibitem[Liang et al.(2008)]{liang08} Liang, N., Xiao, W.~K., Liu, Y., et al.\ 2008, \apj, 685, 354. doi:10.1086/590903
\bibitem[Lusso \& Risaliti(2016)]{lusso16} Lusso, E. \& Risaliti, G.\ 2016, \apj, 819, 154. doi:10.3847/0004-637X/819/2/154
\bibitem[Montiel et al.(2021)]{montiel21} Montiel, A., Cabrera, J.~I., \& Hidalgo, J.~C.\ 2021, \mnras, 501, 3515. doi:10.1093/mnras/staa3926
\bibitem[Mortlock et al.(2011)]{mortlock11} Mortlock, D., Warren, S., Venemans, B., et al.\ 2011, \nat, 474, 616. doi:10.1038/nature10159
\bibitem[Muccino et al.(2021)]{muccino21} Muccino, M., Izzo, L., Luongo, O., et al.\ 2021, \apj, 908, 181. doi:10.3847/1538-4357/abd254
\bibitem[{{Reichart} {et al.} (2001){Reichart},  {Lamb}, {Fenimore } ,{Ramirez-Ruiz}, {Cline} ,
{Hurley}}]{Reichart01}
{Reichart}, D.E., {Lamb}, D.Q., {Fenimore }, E.E., {Ramirez-Ruiz}, E., {Cline} ,T.L.,
{Hurley}, K., 2001, ApJ, 552, 57
\bibitem[Risaliti \& Lusso(2019)]{risaliti19} Risaliti, G. \& Lusso, E.\ 2019, Nature Astronomy, 3, 272. doi:10.1038/s41550-018-0657-z
\bibitem[Salvaterra et al.(2009)]{salvaterra09} Salvaterra, R., Della Valle, M., Campana, S., et al.\ 2009, \nat, 461, 1258. doi:10.1038/nature08445
\bibitem[Sanders et al.(1989)]{sanders89} Sanders, D.~B., Phinney, E.~S., Neugebauer, G., et al.\ 1989, \apj, 347, 29. doi:10.1086/168094
\bibitem[Schaefer(2007)]{schaefer07} Schaefer, B.~E.\ 2007, \apj, 660, 16. doi:10.1086/511742
\bibitem[Shang et al.(2011)]{shang11} Shang, Z., Brotherton, M.~S., Wills, B.~J., et al.\ 2011, \apjs, 196, 2. doi:10.1088/0067-0049/196/1/2
\bibitem[Steffen et al.(2006)]{steffen06} Steffen, A.~T., Strateva, I., Brandt, W.~N., et al.\ 2006, \aj, 131, 2826. doi:10.1086/503627
\bibitem[Strateva et al.(2005)]{strateva05} Strateva, I.~V., Brandt, W.~N., Schneider, D.~P., et al.\ 2005, \aj, 130, 387. doi:10.1086/431247
\bibitem[Tanvir et al.(2009)]{tanvir09} Tanvir, N.~R., Fox, D.~B., Levan, A.~J., et al.\ 2009, \nat, 461, 1254. doi:10.1038/nature08459
\bibitem[Trammell et al.(2007)]{trammell07} Trammell, G.~B., Vanden Berk, D.~E., Schneider, D.~P., et al.\ 2007, \aj, 133, 1780. doi:10.1086/511817
\bibitem[Vignali et al.(2003)]{vignali03} Vignali, C., Brandt, W.~N., \& Schneider, D.~P.\ 2003, \aj, 125, 433. doi:10.1086/345973
\bibitem[Visser(2004)]{visser04} Visser, M.\ 2004, Classical and Quantum Gravity, 21, 2603. doi:10.1088/0264-9381/21/11/006
\bibitem[Vitagliano et al.(2010)]{vitagliano10} Vitagliano, V., Xia, J.-Q., Liberati, S., et al.\ 2010, \jcap, 2010, 005. doi:10.1088/1475-7516/2010/03/005
\bibitem[Wang et al.(2007)]{wang07} Wang, F.~Y., Dai, Z.~G., \& Zhu, Z.-H.\ 2007, \apj, 667, 1. doi:10.1086/52076
\bibitem[Wang et al.(2015)]{wang15} Wang, F.~Y., Dai, Z.~G., \& Liang, E.~W.\ 2015, \nar, 67, 1. doi:10.1016/j.newar.2015.03.001
\bibitem[Wang et al.(2017)]{wang17} Wang, G.-J., Yu, H., Li, Z.-X., et al.\ 2017, \apj, 836, 103. doi:10.3847/1538-4357/aa5b9b
\bibitem[Wang et al.(2018)]{wang18} Wang, F.-G., Yang, J.-Y., Fan, X.-H., et al.\ 2018, \apj, 869, L9. doi:10.3847/2041-8213/aaf1d2
\bibitem[Wei et al.(2013)]{wei13} Wei, J.-J., Wu, X.-F., \& Melia, F.\ 2013, \apj, 772, 43. doi:10.1088/0004-637X/772/1/43
\bibitem[Weinberg(1972)]{weinberg72} Weinberg, S.\ 1972, Gravitation and Cosmology: Principles and Applications of the General Theory of Relativity, by Steven Weinberg, pp. 688. ISBN 0-471-92567-5. Wiley-VCH , July 1972., 688
\bibitem[Yang et al.(2020)]{yang20} Yang, J.-Y., Wang, F.-G., Fan, X.-H., et al.\ 2020, \apj, 897, L14. doi:10.3847/2041-8213/ab9c26
\bibitem[Young et al.(2010)]{young10} Young, M., Elvis, M., \& Risaliti, G.\ 2010, \apj, 708, 1388. doi:10.1088/0004-637X/708/2/1388
\bibitem[Zhang(2018)]{zhang18} Zhang, B.\ 2018, The Physics of Gamma-Ray Bursts by Bing Zhang. ISBN: 978-1-139-22653-0. Cambridge Univeristy Press, 2018. doi:10.1017/9781139226530
\bibitem[Zheng et al.(2020)]{zheng20} Zheng, X.-G., Liao, K., Marek, B., et al.\ 2020, \apj, 892, 103. doi:10.3847/1538-4357/ab7995
\bibitem[Zheng et al.(2021)]{zheng21} Zheng, X.-G., Cao, S., Biesiada, M., et al.\ 2021, arXiv: 2103.07139, doi: 10.1007/s11433-020-1664-9

\end{thebibliography}
\end{document}